\documentclass[twocolumn,showpacs,pra]{revtex4}

\usepackage{graphicx}

\begin{document}

\title{Simulations of ultracold bosonic atoms in optical lattices with 
anharmonic traps}

\author{Olivier Gygi}
\author{Helmut G.~Katzgraber}
\author{Matthias Troyer}
\affiliation{Theoretische Physik, ETH Z\"urich, CH-8093 Z\"urich, 
Switzerland}

\author{Stefan Wessel} 
\affiliation{Institut f\"ur Theoretische Physik III, Universit\"at 
Stuttgart, 70550 Stuttgart, Germany}
\author{G.~George Batrouni}
\affiliation{Institut Non-Lin\'eaire de Nice, UMR CNRS 6618,
  Universit{\'e} de Nice-Sophia Antipolis, 
1361 route des Lucioles, F--06560 Valbonne, France}

\date{\today}

\begin{abstract}
We report results of quantum Monte Carlo simulations in the canonical
and the grand-canonical ensemble of the two- and three-dimensional
Bose-Hubbard model with quadratic and quartic confining
potentials. The quantum criticality of the superfluid--Mott insulator
transition is investigated both on the boundary layer separating the
two coexisting phases and at the center of the traps where the
Mott-insulating phase is first established. Recent simulations of
systems in quadratic traps have shown that the transition is not in
the critical regime due to the finite gradient of the confining
potential and that critical fluctuations are suppressed. In addition,
it has been shown that quantum critical behavior is recovered in flat
confining potentials as they approach the uniform regime. Our results
show that quartic traps display a behavior similar to
quadratic ones, yet locally at the center of the traps the bulk
transition has enhanced critical fluctuations in comparison to the 
quadratic case. Therefore quartic traps
provide a better prerequisite for the experimental observation of true
quantum criticality of ultracold bosonic atoms in optical lattices.
\end{abstract}

\maketitle

\section{Introduction}
\label{sec:introduction}

The Bose-Hubbard model offers an almost perfect description of
Bose-Einstein condensates in optical
lattices~\cite{jaksch:98,greiner:02}. Therefore the model has been intensively
studied both
analytically~\cite{fisher:89,niyaz:94,jaksch:98,zwerger:03} as well as
numerically~\cite{batrouni:90,batrouni:96,freericks:96,kashurnikov:02,schmid:02,batrouni:02,wessel:04,wessel:05}
in recent years. In the absence of a trapping potential, i.e., for the
{\em homogeneous Bose-Hubbard model}, a quantum phase transition from
a superfluid to a Mott-insulating phase occurs at commensurate
fillings upon increasing the optical lattice
depth~\cite{fisher:89}. Experimentally, Bose-Einstein condensates are
created by cooling bosonic atoms. In order to prevent them from
evaporating, they must be trapped by a confining potential. As shown
for example in Ref.~\onlinecite{batrouni:02}, inhomogeneous confining
potentials induce a coexistence of superfluid and Mott-insulating
regions for a continuous range of incommensurate
fillings. By a suitable increase of the lattice depth and the number of 
particles, one may start forming a Mott-insulating region in the center of 
the trap. A further increase of the lattice depth induces a change of the 
volume fractions between the two phases. Accordingly, the transition recently
observed in experiments of ultracold Bose gases in optical lattices
embedded in quadratic confining potentials~\cite{greiner:02} should be
better viewed as a crossover with changing volume fractions of the two
phases~\cite{wessel:04} rather than as a true phase transition.

Although globally no true transition can be observed in systems with
inhomogeneous traps, one might expect to observe a superfluid--Mott
insulator transition on the boundary layer separating the coexisting
superfluid and Mott-insulating regions and, locally, at the center of
the trap where the Mott-insulating phase has been established.  In the
following we distinguish these transitions by referring to them as
``surface transition'' and ``bulk transition,'' respectively.  Recent
quantum Monte Carlo simulations of the Bose-Hubbard model in quadratic
confining potentials have shown that the quantum criticality of these
transitions is destroyed by the finite gradient of the confining
potential~\cite{wessel:04}. However it has been demonstrated that
quantum criticality is recovered in flat traps where the potential
gradient becomes irrelevant in the center, i.e., when locally the trap
center approaches the uniform regime.

Here we study flat anharmonic traps, which can be experimentally
realized, e.g., by superimposing pairs of weak, repulsive,
blue-detuned Gaussian laser beams to pairs of attractive, red-detuned
Gaussian laser beams that create the optical lattice. Depending on the
beam parameters, it is possible to cancel out the harmonic terms in
the series expansion of the resulting potential. In our simulations
with anharmonic traps we consider only fourth-order terms in the
expansion and neglect all higher-order ones.  The results we have
obtained from our simulations of two-dimensional (2D) systems in
quartic traps and the comparison to our results for systems in
quadratic traps, closed-box systems confined by completely flat and
infinitely sharp traps, as well as homogeneous systems with periodic
boundary conditions clearly demonstrate that the surface transition in
systems with quartic traps has the same lack of quantum criticality as
the corresponding transition in the systems with quadratic traps.
However, the bulk transition has stronger critical fluctuations in the
center of quartic traps, than in the center of quadratic traps.
These observations lead us to the conclusion that quantum
criticality in systems with inhomogeneous traps is determined by the
flatness in the trap centers and their closest surroundings, i.e., by
the smaller deviation from the uniform regime, and not by the shape
and steepness of the walls.  Thus, quartic traps could provide a
better prerequisite for future experimental observations of true
quantum criticality.

We also present results on three-dimensional (3D) systems with smaller system
sizes which qualitatively show the same results as in two space dimensions, 
albeit less pronounced.

The paper is organized as follows: In Sec.~\ref{sec:model} we introduce in 
detail both the Bose-Hubbard model with the specific traps that we have 
used and the observables needed for the investigations of the systems. In 
Sec.~\ref{sec:results} we present our results, and in 
Sec.~\ref{sec:conclusions} we summarize our observations.

\section{Model, Observables, and Numerical Details}
\label{sec:model}

In order to describe trapped ultracold bosonic atoms in optical 
lattices we use the Bose-Hubbard Hamiltonian
\begin{eqnarray}
{\mathcal{H}} = & - & t\sum_{\langle i,j\rangle}
\left(\hat{b}^{\dagger}_i\hat{b}_j+\mathrm{H.c.}\right)
+\frac{U}{2}\sum_i\hat{n}_i(\hat{n}_i-1)\nonumber\\
& - & \mu_0\sum_i\hat{n}_i+\sum_iV(i)\hat{n}_i,
\label{eq:hamiltonian}
\end{eqnarray}
where $\hat{b}^{\dagger}_i$ and $\hat{b}_i$ are the creation and 
annihilation operators for bosons at lattice sites $i$, respectively, and 
\begin{equation}
\hat{n}_i=\hat{b}^{\dagger}_i\hat{b}_i
\label{eq:local_density_operator}
\end{equation}
is the local density operator. $t$ represents the nearest-neighbor hopping 
matrix element, $U$ the on-site repulsion, $\mu_0$ a chemical potential 
offset that controls the filling of the trap and $V(i)$ the trap potential. The two last terms of the Hamiltonian can be conveniently combined by defining 
an effective, spatially-dependent chemical potential $\mu_i^{\mathrm{eff}}$ 
that is experienced  by a boson at site $i$, i.e., 
\begin{equation}
\mu_{i}^{\mathrm{eff}}=\mu_0-V(i).
\label{eq:mueff}
\end{equation}

Due to the inhomogeneity introduced by a trapping potential needed for the 
confinement of the atoms in experimental realizations of the Bose-Hubbard 
model, it becomes necessary to consider local, site-dependent quantities, 
in contrast to the homogeneous Bose-Hubbard model, where it suffices to 
measure global quantities such as the global compressibility and the 
superfluid density~\cite{pollock:87} to capture the system 
characteristics. One way of distinguishing coexisting local superfluid and 
Mott-insulating states in systems with traps is to investigate the 
topology of local density profiles. Regions with constant integer fillings 
can be interpreted to be Mott insulating, whereas regions with noninteger 
fillings must be superfluid. The local compressibility 
$\kappa_{i}^{\mathrm{local}}$ at site $i$
\begin{equation}
\kappa_{i}^{\mathrm{local}} =
\frac{\partial\langle N\rangle}{\partial \mu_i^{\mathrm{eff}}} =
\int_0^{\beta}\mathrm{d}\tau
\left[\langle \hat{n}_i(\tau)N\rangle-\langle\hat{n}_i(\tau)\rangle\langle N\rangle\right],
\label{eq:kappalocal}
\end{equation}
provides a more precise way to distinguish the local states and to probe 
for quantum criticality. It expresses the response of the system's 
particle number $N$ to a local change of the effective chemical potential 
$\mu_i^{\mathrm{eff}}$ at site $i$. In the above equation, $\beta=1/k_BT$ 
denotes the inverse temperature
\begin{equation}
\hat{n}_i(\tau)=\exp(\tau{\mathcal H})n_i\exp(-\tau{\mathcal H}) ;
\label{eq:propagator}
\end{equation}
the imaginary-time propagated operator, and $\langle\cdots\rangle$ the 
Monte Carlo sample average. Note that this definition of the local 
compressibility only makes sense in the grand-canonical ensemble where the 
total number of particles $N$ in the system is variable, since otherwise 
the integrand vanishes and the local compressibility is zero everywhere.

The measurement of local density fluctuations provides another way of
investigating the local states of inhomogeneous systems. In addition
to measuring the variance of the local density
\begin{equation}
\Delta_i=\langle n_i^2\rangle-\langle n_i\rangle^2,
\label{eq:variance}
\end{equation}
it can be useful to measure the on-site compressibility
$\kappa_{i}^{\mathrm{onsite}}$, i.e., the response 
of the local density at site $i$ to a chemical potential change at this 
site
\begin{equation}
\kappa_{i}^{\mathrm{onsite}} =
\frac{\partial\langle n_i\rangle}{\partial \mu_i^{\mathrm{eff}}} =
\int_0^{\beta}\mathrm{d}\tau\left[\langle \hat{n}_i(\tau)\hat{n}_i(0)\rangle-\langle\hat{n}_i(\tau)\rangle\langle \hat{n}_i(0)\rangle\right].
\label{eq:kappaonsite}
\end{equation}

\begin{figure}
\includegraphics[width=\columnwidth]{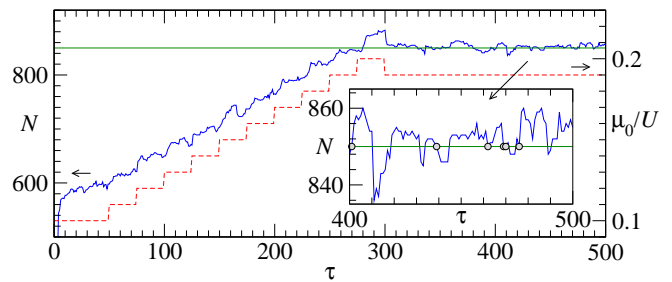}
\caption{(Color online) Evolution of the measured particle number $N$
(solid curve) toward the desired value $N=850$ while adjusting the
chemical potential offset $\mu_0/U$ (dashed curve) at the beginning of the
thermalization phase. The system size is $50\times50$ in a quartic trap.
The inset is a detailed view of $N$ fluctuating around $N=850$ as a
function of time. The circles indicate the configurations for which
$N$ is exactly $850$ and which can be used in a canonical measurement.
}
\label{fig:adjustment}
\end{figure}

For our simulations, we have used two wormlike quantum Monte Carlo 
algorithms provided by the ALPS project~\cite{alet:05b}, the worm 
algorithm~\cite{prokof'ev:98}, and the stochastic series expansion (SSE) 
algorithm with directed loops~\cite{alet:05a}. Both algorithms work 
generically in the grand-canonical ensemble. 
Since experiments are usually performed with a fixed number of particles, 
we want our simulations to also be performed with a fixed particle number, 
i.e., in the canonical ensemble. For this purpose, we have applied some 
modifications to the algorithms to fit our needs. In order for the 
Hamiltonian in Eq.~(\ref{eq:hamiltonian}) to be fully defined, the correct 
chemical potential offset $\mu_0/U$, which yields the desired particle 
number $N$ at which the simulations are to be performed, has to be 
determined. Starting from a guess value the correct 
value of $\mu_0/U$ is approximated at the beginning of the thermalization 
phase of the simulation, as illustrated in Fig.~\ref{fig:adjustment}. The 
dashed curve shows the subsequent adjustments of the chemical potential 
offset $\mu_0/U$ and the solid curve represents the evolution of the 
measured particle number, which equilibrates around the desired particle 
number $N$ (solid, horizontal line) once the adjustment of $\mu_0/U$ is 
complete.

Measuring observables in the canonical ensemble is then simply done by 
measuring the particle number after each Monte Carlo step and by selecting 
out only updated configurations for which the measured particle number 
coincides with the desired particle number at which the simulations are 
meant to be performed. By computing the Monte Carlo sample averages using 
only the selected updated configurations, the observables are evaluated in 
the canonical ensemble, whereas by including all other updated 
configurations for the computation of the sample averages, the observables 
are evaluated in the grand-canonical ensemble.

We have found no substantial qualitative deviations between the canonical 
and grand-canonical data for all observables. In Figs.~\ref{fig:2D_1} 
and~\ref{fig:2D_2} we plot the data evaluated in the canonical ensemble, 
except for the local compressibility, which, for reasons mentioned above, 
cannot be measured canonically. In Figs.~\ref{fig:2D_3},~\ref{fig:2D_5} 
and~\ref{fig:3D_1}, where comparisons are made with uniform systems, we 
have chosen to plot the data evaluated in the grand-canonical ensemble, 
since obtaining the different curves for the uniform systems implies 
that the filling is allowed to change \cite{mu_scan}.

\begin{table}
\caption{
Overview of the simulations with quadratic and quartic traps in 2D simulated
in the canonical ensemble. $N$ is the fixed particle number at which the 
observables are evaluated in the canonical ensemble, $\mu_0/U$ and $t/U$ 
are the chemical potential offset and the hopping parameter in units of the 
interaction parameter $U$, respectively. $a_2$ and $a_4$ are the curvatures 
of the quadratic and quartic traps. 
}
\label{tab:parameters_c}
\begin{tabular*}{\columnwidth}{@{\extracolsep{\fill}} l l c c c }
\hline
\hline
Size & Trap curvature & $N$ & $\mu_0/U$ & $t/U$\\
\hline
$50^2$ & $a_4/U=3.5\cdot10^{-6}$ & $1200$ & $0.362$ & $0.04$\\
$50^2$ & $a_2/U=2\cdot10^{-3}$ & $600$ & $0.370$ & $0.04$\\
\hline
\hline
\end{tabular*}
\end{table}%

\begin{table}
\caption{
Overview of the simulations with quadratic and quartic traps in 2D (top) and 
3D (bottom) simulated in the grand-canonical ensemble. $N$ is the particle 
number and $\mu^{\mathrm{eff}}_{\mathrm{center}}/U$ and $t/U$ are the 
effective chemical potential at the central lattice sites~\cite{mu_dilemma} 
and the hopping parameter in units of the interaction parameter $U$, 
respectively. $a_2$ and $a_4$ are the curvatures of the quadratic and 
quartic traps. 
}
\label{tab:parameters_gc}
\begin{tabular*}{\columnwidth}{@{\extracolsep{\fill}} l l c c c }

\hline
\hline
Size & Trap curvature & $N$ & $\mu^{\mathrm{eff}}_{\mathrm{center}}/U$ & $t/U$\\
\hline
$50^2$ & $a_4/U=3.5\cdot10^{-6}$ & $908$ -- $924$ & $0.217$ & $0.01$ -- $0.06$\\
$50^2$ & $a_2/U=2\cdot10^{-3}$   & $341$ -- $413$ & $0.217$ & $0.01$ -- $0.06$\\
\hline
$14^3$ & $a_4/U=7\cdot10^{-4}$   & $462$ -- $498$ & $0.217$ & $0.006$ -- $0.04$\\
$14^3$ & $a_2/U=3\cdot10^{-2}$   & $90$  -- $122$ & $0.217$ & $0.006$ -- $0.04$\\
\hline
\hline
\end{tabular*}
\end{table}%

\begin{figure}
\includegraphics[width=\columnwidth]{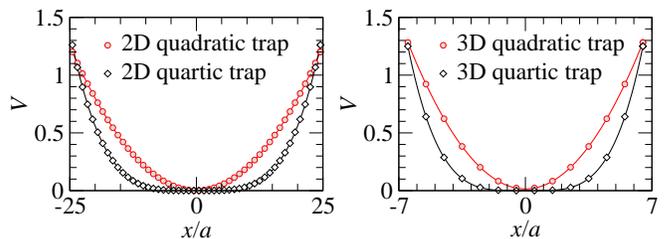}
\caption{(Color online)
Sections  $V(x,y=a/2)$ and $V(x,y=a/2,z=a/2)$ through 
the 2D (left) and 3D (right) traps that have been used in the simulations. 
The simulation parameters are summarized in Tables ~\ref{tab:parameters_c} and 
\ref{tab:parameters_gc}. The points indicate the lattice sites
separated by the lattice constant $a$.
}
\label{fig:traps}
\end{figure}

The 2D and 3D trapping potentials we have used are
\begin{eqnarray}
V^{(2)}_{\mathrm{2D}} & = & a_2\left(x^2+y^2\right) ,\\
V^{(4)}_{\mathrm{2D}} & = & a_4\left(x^4+y^4\right) ,\\
V^{(2)}_{\mathrm{3D}} & = & a_2\left(x^2+y^2+z^2\right) ,\\ 
V^{(4)}_{\mathrm{3D}} & = & a_4\left(x^4+y^4+z^4\right) ,
\end{eqnarray}
where the trap curvatures are given in Tables~\ref{tab:parameters_c}
and~\ref{tab:parameters_gc} (see also Fig.~\ref{fig:traps}).  The
lattice sites in the 2D and 3D systems in units of the lattice spacing
$a$ are located at the coordinates $(x,y) = (\pm[n+1/2],\pm[n+1/2])$
and $(x,y,z) = (\pm[n+1/2],\pm[n+1/2],\pm[n+1/2])$, respectively,
where $n=0,1,\dots,L/2-1$. The linear extents of the systems
with quadratic and quartic traps are $L/a=50$ in 2D and $L/a=14$ in
3D. With the trap curvatures and the range of the chemical potential
offsets $\mu_0/U$ that we have used, these system sizes ensure that
the effective chemical potential takes large enough negative values at
the boundaries of the systems for the particles to never reach the
boundaries. The linear extents of the systems with closed boxes and
the homogeneous systems with periodic boundary conditions have been
reduced to $L/a=32$ and $L/a=28$ in 2D and $L/a=8$ in 3D to
approximatively match the effective number of bosons trapped in the
quadratic and quartic traps, i.e., the sizes of the regions with a
nonvanishing local density.

\section{Results}
\label{sec:results}

In this section we present the results obtained from simulations of the 
two- and three-dimensional Bose-Hubbard model in quartic traps and 
compare results for quadratic traps with closed box configurations 
and homogeneous systems with periodic boundary conditions. 
Tables~\ref{tab:parameters_c} and \ref{tab:parameters_gc} give an overview 
of the simulation parameters in the canonical and grand-canonical ensemble,
respectively.

First, we investigate quantum criticality at the {\em surface} which separates 
the superfluid and Mott-insulating states, which is shell structured
in systems with quadratic and quartic traps. Our results suggest that in
regards to quantum criticality at the surface quadratic and quartic traps 
are equivalent: both systems show the same lack of quantum criticality.
Next, we study quantum criticality to the {\em center} of the traps
by restricting the measurements to small regions around the center.
We drive these regions through the superfluid--Mott insulator 
transition and show that this {\em bulk} transition in the center of
quartic traps has enhanced critical fluctuations in comparison to
the center of quadratic traps. 

\subsection{Quantum criticality on the boundary between superfluid and
  Mott-insulating regions} 
\label{sec:qc_coexisting}

\begin{figure}
\includegraphics[width=4cm]{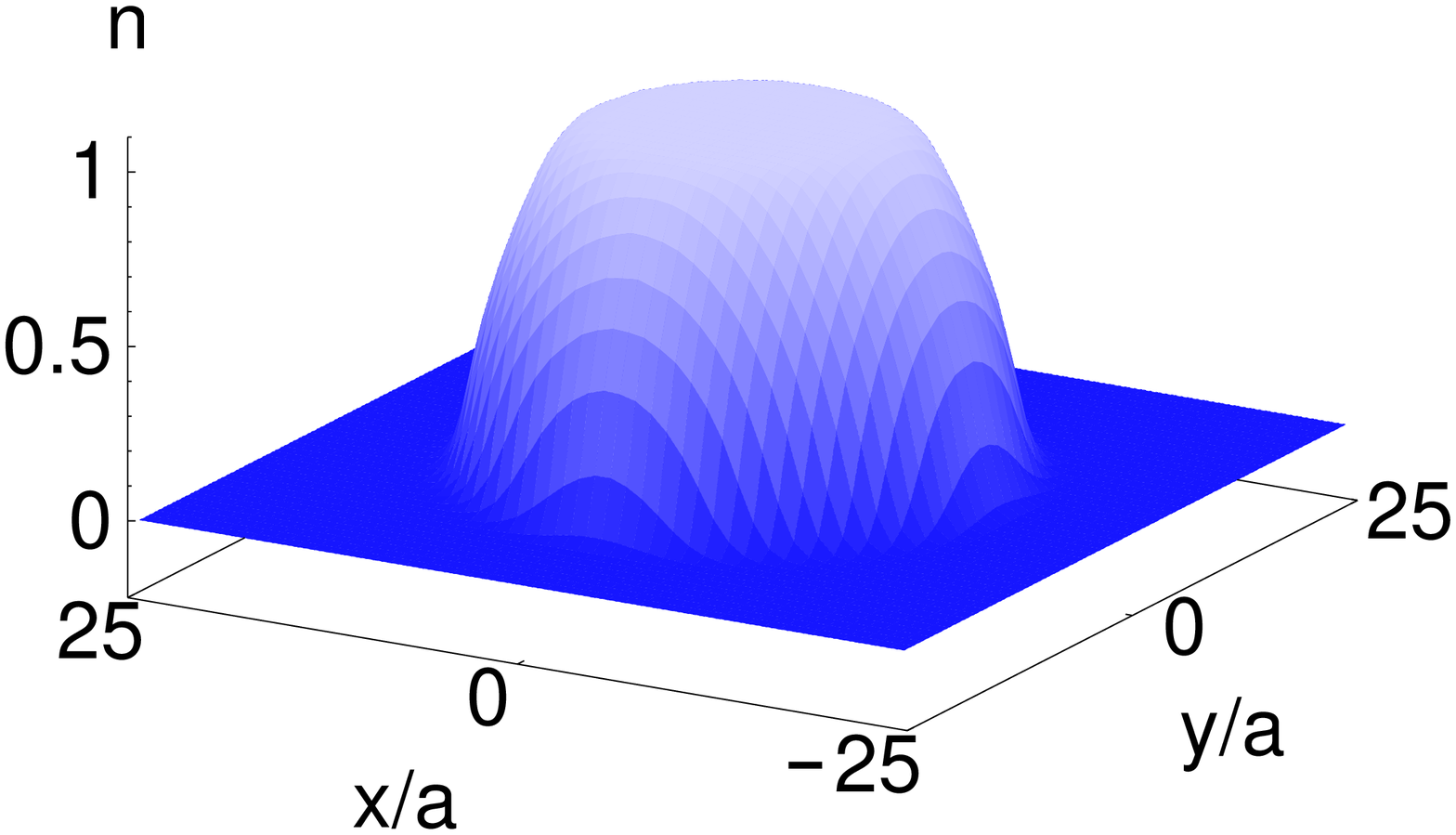}
\includegraphics[width=4cm]{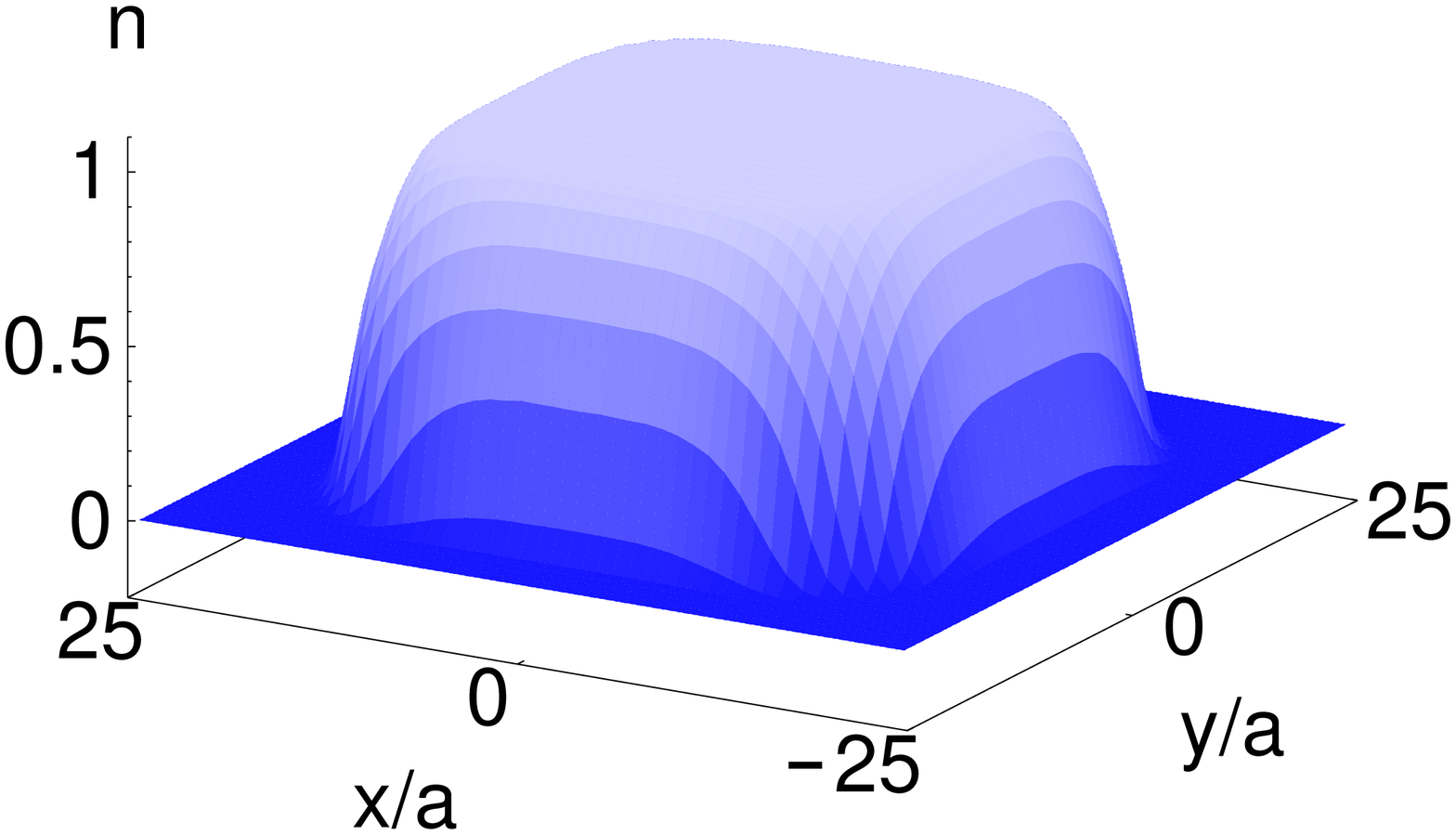}

\includegraphics[width=4cm]{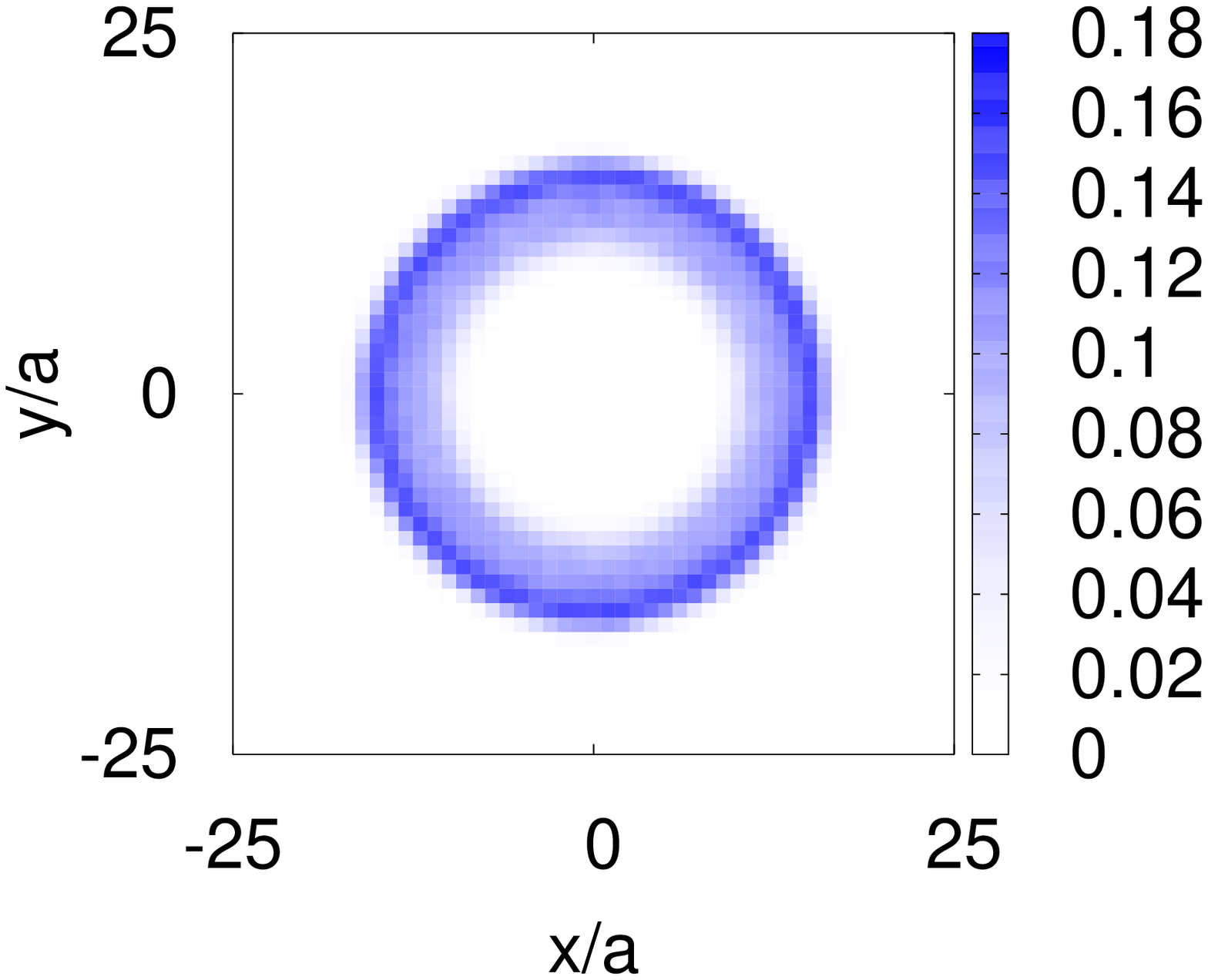}
\includegraphics[width=4cm]{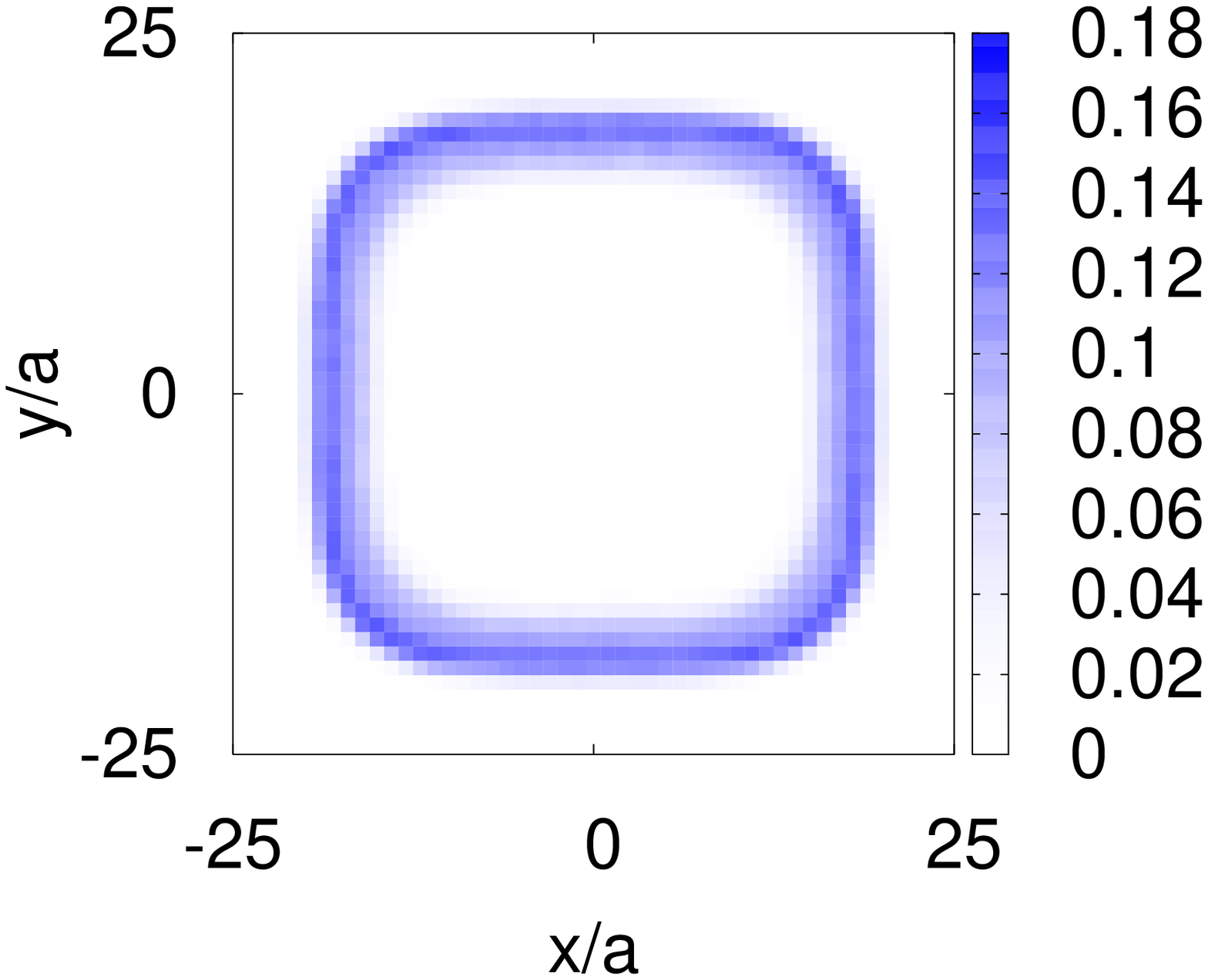}
\caption{(Color online)
Spatial dependence of the local density $n$ measured in the canonical 
ensemble (top), and of the local compressibility $\kappa^{\mathrm{local}} t$ 
measured in the grand-canonical ensemble (bottom) for 2D systems of size 
$50\times50$ containing respectively 600 bosons in a quadratic trap with 
curvature $a_2/U=2\cdot10^{-3}$, chemical potential offset $\mu_0/U=0.370$ 
and hopping parameter $t/U=0.04$ (left) and 1200 bosons in a quartic trap 
with curvature $a_4/U=3.5\cdot10^{-6}$, chemical potential offset 
$\mu_0/U=0.362$ and the same hopping parameter $t/U=0.04$ (right). The 
special fourfold symmetry of the quartic trap is clearly reflected by the 
shape of the superfluid shell between the outer $n=0$ and the central 
$n=1$ Mott plateaus (right).
}
\label{fig:2D_1}
\end{figure}

We compare a system with a quadratic trap containing 600 bosons and 
a system with a quartic trap containing 1200 bosons. The hopping parameter 
is $t/U=0.04$ in both cases and the chemical potential offsets are 
$\mu_0/U=0.370$ in the quadratic case and $\mu_0/U=0.362$ in the quartic 
case. In both cases a Mott-insulating plateau with integer density 
$\langle n_i\rangle=1$ is present at the center of the trap, which is 
surrounded by a superfluid ringlike region, as depicted in 
Fig.~\ref{fig:2D_1}, where the spatial dependence of the local density $n$ 
and of the local compressibility $\kappa^{\mathrm{local}}$, is shown. The 
left panels show the quadratic case and the right panels the quartic case. 
In both cases, the local compressibility takes its largest values close to 
the outer boundary of the superfluid shell. The shape of the superfluid 
shell clearly reflects the special fourfold rotational symmetry of the 
quartic trap, compared to the quadratic trap, which has a continuous 
rotational symmetry. Furthermore, the width of the superfluid shell is 
clearly smaller in the quartic trap than in the quadratic trap, which is 
due to the larger steepness of the former in the corresponding regions of 
the superfluid shell.

\begin{figure}
\includegraphics[width=\columnwidth]{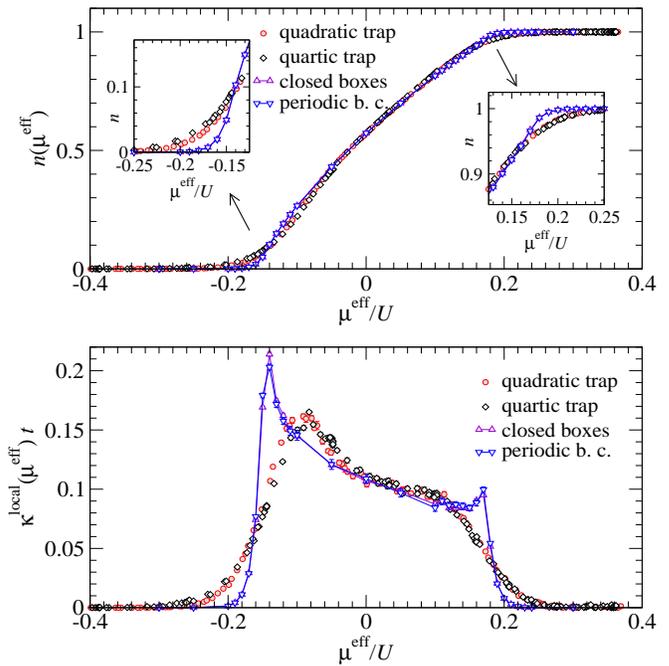}
\caption{(Color online) Local density $n$ (top) and local
compressibility $\kappa^{\mathrm{local}}$ (bottom), measured in the
grand-canonical ensemble, vs the local effective chemical potential
$\mu^{\mathrm{eff}}/U$. The quadratic and quartic trap cases are the
same as Fig.~\ref{fig:2D_1}. We also show systems in closed boxes of
size $32\times32$ as well as homogeneous systems with periodic
boundary conditions also of size $32\times32$. In all system, the
hopping parameter is $t/U=0.04$. }
\label{fig:2D_3}
\end{figure}

A better quantitative description of the systems in the two different
traps is given by the radial dependence of the observables from the
center towards the boundaries of the systems. 
In the quadratic case, due to the continuous rotational
symmetry, the values at all lattice sites can be used to
create such geometric profiles. In the quartic case, we are limited by
the fourfold symmetry of the underlying potential and we thus
use only points lying on the sections through the center and
parallel to the boundaries of the systems.
This limitation no longer applies to quartic traps when we plot
local quantities as a function of the local effective chemical potential
$\mu^{\mathrm{eff}}/U$, since, as follows from the data collapse
on single curves, cf.~Fig.~\ref{fig:2D_3}, a local potential
approximation holds just as for quadratic traps~\cite{wessel:04},
i.e., these quantities can be determined from the value of the local
effective chemical potential.

The errors have been determined by means of a nonparametric bootstrap 
analysis~\cite{efron:79} applied to each set of values corresponding to 
the same radial distance from the center or to the same effective chemical 
potential. If the error bars are not visible, they are smaller than the 
corresponding point sizes.

\begin{figure}
\includegraphics[width=\columnwidth]{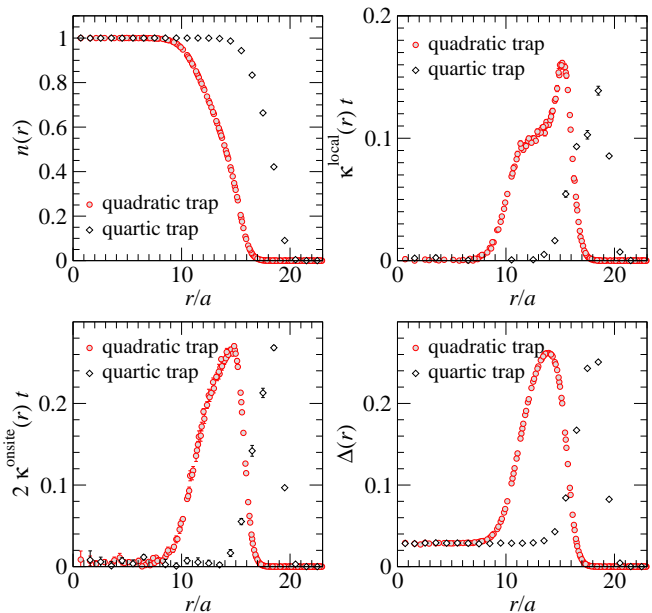}
\caption{(Color online)
Radial dependence of the local density $n$, the local compressibility 
$\kappa^{\mathrm{local}}$, the on-site compressibility 
$\kappa^{\mathrm{onsite}}$ and the variance $\Delta$ for the same systems 
as in Fig.~\ref{fig:2D_1}. All quantities are measured in the canonical 
ensemble, except for $\kappa^{\mathrm{local}}$ that can only be measured 
in the grand-canonical ensemble.
}
\label{fig:2D_2}
\end{figure}

The radial dependence of the local density $n$, the local
compressibility $\kappa^{\mathrm{local}}$, the on-site compressibility
$\kappa^{\mathrm{onsite}}$ and the variance $\Delta$ for the same
systems introduced above are given in Fig.~\ref{fig:2D_2}. Noninteger
local densities correspond to regions where the system is in the
superfluid state, whereas integer local densities indicate regions in
which the system is in the Mott-insulating state. The local
compressibility profile consists of an asymmetric double peak, which
reflects the increase of the particle number fluctuations near the
boundaries of the Mott-insulating regions. Both peaks are of the same
height in the corresponding hard-core model due to particle-hole
symmetry~\cite{wessel:04}. The on-site compressibility,
$\kappa^{\mathrm{onsite}}$, and the variance, $\Delta$, both peak
inside the superfluid shell, but in contrast to the local
compressibility, $\kappa^{\mathrm{local}}$, they do not completely
vanish inside the central Mott plateau. This is due to virtual hopping
processes that are completely suppressed only by an infinite energy
gap in the limit $t/U\rightarrow0$~\cite{wessel:04}. As for the
quadratic traps~\cite{wessel:04}, we find for the quartic traps that
$\kappa^{\mathrm{local}}$ is a better probe than
$\kappa^{\mathrm{onsite}}$ and $\Delta$ for the existence of
superfluid and Mott-insulating regions. It can, therefore, serve as a
genuine order parameter to characterize the superfluid--Mott insulator
transition.

In Fig.~\ref{fig:2D_3} we show the local density $n$ and the local
compressibility $\kappa^{\mathrm{local}}$ as functions of the local
effective chemical potential $\mu^{\mathrm{eff}}/U$, which, as pointed
out in Ref.~\onlinecite{wessel:04}, are not universal functions, but
depend on both the geometries and the curvatures of the traps. The
data for both quantities taken from the simulations with quadratic and
quartic traps match almost perfectly. As a comparison, both curves are
plotted for closed-box systems confined by completely flat and
infinitely sharp traps as well as for homogeneous systems with
periodic boundary conditions, all of size $32\times32$. This smaller
size is comparable to the effective sizes of the systems with the
quadratic and the quartic traps. Since in these uniform cases the
effective chemical potential remains constant over the whole system, a
whole set of simulations, each of which is performed with a different
chemical potential, is needed in order to obtain these additional
curves.  In the local density curve, cusps appear at the points where
the density approaches $n=0$ and $n=1$ and correspondingly both
singularities in the local compressibility are more pronounced,
indicating quantum criticality. As shown in
Ref.~\onlinecite{wessel:04}, these cusps in $n(\mu^{\mathrm{eff}}/U)$
become smoother as the curvature is increased from zero (uniform case)
to finite, positive values. This is due to the gradient in the
confining potential that becomes increasingly relevant and which
destroys the quantum criticality.  No quantum critical behavior is
found for the transition at the surface separating the coexisting
superfluid and Mott-insulating states in the quartic trap, in spite of
its flatness in the center and the larger steepness of its walls.

\subsection{Bulk quantum criticality in the trap center}
\label{sec:qc_center}

\begin{figure}
\includegraphics[width=\columnwidth]{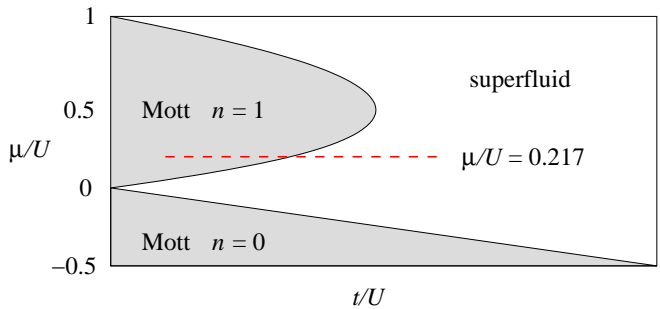}
\caption{(Color online)
Phase diagram of the homogeneous Bose-Hubbard model~\cite{fisher:89}. It 
consists of a series of Mott-insulating lobes that are surrounded by a 
superfluid region. Each of the Mott lobes is characterized by an integer 
filling of single lattice sites. Here, only the $n=0$ and $n=1$ Mott lobes 
are shown. The constant $(\mu/U=0.217)$ scans presented in 
Figs.~\ref{fig:2D_5} and~\ref{fig:3D_1} are indicated by the dashed, 
horizontal line.
}
\label{fig:phase_diagram}
\end{figure}

\begin{figure}
\includegraphics[width=\columnwidth]{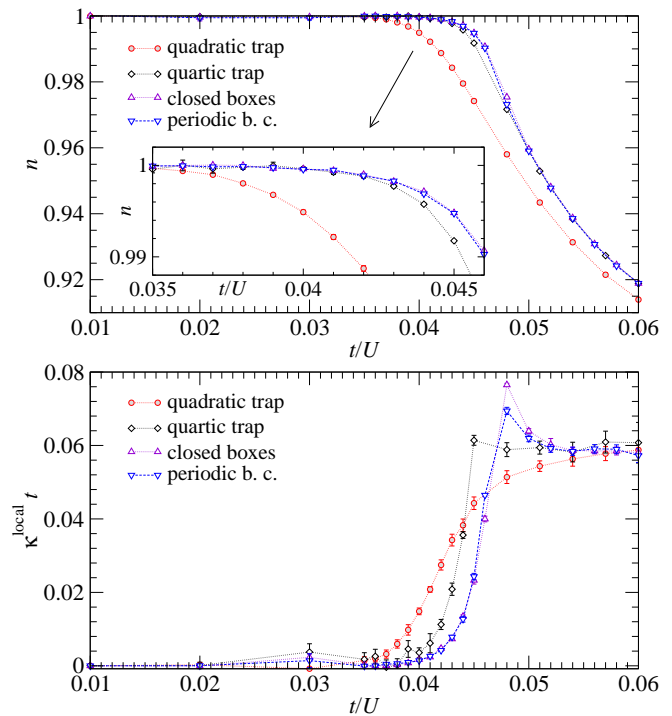}
\caption{(Color online) Constant $(\mu/U=0.217)$ scans through the
critical point. The density $n$ (top) and the local compressibility
$\kappa^{\mathrm{local}}$ (bottom), in the grand-canonical ensemble,
vs $t/U$ for 2D systems of $50\times50$ sites. The quadratic trap
curvature is $a_2/U=2\cdot10^{-3}$ and the quartic trap curvature
$a_4/U=3.5\cdot10^{-6}$. Both quantities are measured at the center of
the traps. For comparison, the same scans are plotted for 
a system in a closed-box and one with periodic boundary conditions of
size $28\times28$.
}
\label{fig:2D_5}
\end{figure}

Instead of investigating the transition at the surface separating the 
coexisting superfluid and Mott-insulating states in the systems with 
quadratic and quartic traps, we now focus on the center of these traps and 
drive the local state through the superfluid--Mott insulator transition, 
which can be induced by either increasing the chemical potential 
offset $\mu_0/U$ or by decreasing 
the ratio $t/U$ between the hopping and the interaction 
parameter. The latter case, which is more likely to be reproduced 
experimentally, is shown in Fig.~\ref{fig:2D_5}, where the local density 
$n$ and the local compressibility $\kappa^{\mathrm{local}}$ are plotted 
against $t/U$ at constant chemical potential $\mu/U=0.217$. This scan is 
marked with a dashed line in Fig.~\ref{fig:phase_diagram} where the phase 
diagram of the homogeneous Bose-Hubbard model~\cite{fisher:89} is depicted 
schematically. At each point of the scan, both $n$ and 
$\kappa^{\mathrm{local}}$ are measured at the twelve lattice sites that 
are closest to the center, i.e., at the four central lattice 
sites~\cite{mu_dilemma} and at their eight nearest neighbors (cross
configuration). In 2D, with our chosen system sizes and trap curvatures, 
the difference of the effective chemical potential felt by bosons on the 
four central lattice sites and the one felt by bosons on their nearest 
neighbor lattice sites is negligible and thus the values measured at the 
nearest neighbor lattice sites can be included to enhance the statistics 
without introducing any bias. The curves originating from the system with 
the quartic trap almost match the curves originating from the uniform 
systems with closed boxes and periodic boundary conditions, whereas for the 
curves obtained from the system with the quadratic trap, the deviations 
are much stronger. The sharp cusp in the $\kappa^{\mathrm{local}}$ curve 
of the system with the quartic trap indicates the increase of critical 
fluctuations inside the trap and will diverge in the continuum limit.
This special behavior demonstrates that locally at the center, the bulk 
transition from the superfluid to the Mott-insulating state 
exhibits stronger critical fluctuations in the quartic trap than in the 
quadratic trap.

\begin{figure}
\includegraphics[width=\columnwidth]{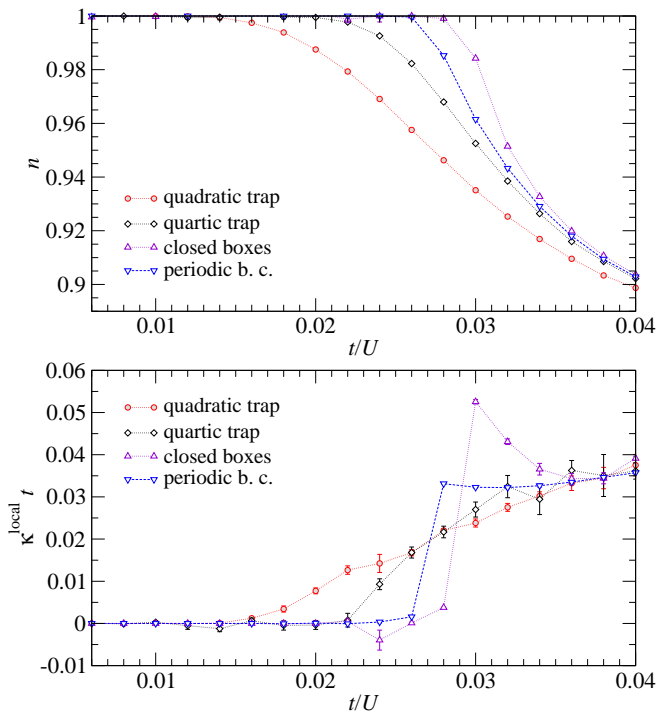}
\caption{(Color online) Constant $(\mu/U=0.217)$ scans through the
critical point. The density $n$ (top) and the local compressibility
$\kappa^{\mathrm{local}}$ (bottom), in the grand-canonical ensemble,
are plotted for 3D systems of $14^3$ sites in a quadratic trap
(curvature $a_2/U=0.03$) and in a quartic trap (curvature
$a_4/U=7\cdot10^{-4}$). As in the 2D scans (Fig.~\ref{fig:2D_5}), $n$
and $\kappa^{\mathrm{local}}$ are measured at the center of the
traps. For comparison, the same scans are plotted for closed-box
systems and homogeneous systems with periodic boundary conditions of
size $8^3$, which approximatively corresponds to the effective sizes
of the systems with the quadratic and quartic traps.
}
\label{fig:3D_1}
\end{figure}

Similarly as in 2D, we have performed the same constant ($\mu/U=0.217$) 
scans through the critical point also for 3D systems, see 
Fig.~\ref{fig:phase_diagram}. In contrast to the 2D case, the 
local density $n$ and the local compressibility $\kappa^{\mathrm{local}}$ 
have been measured only at the eight central lattice 
sites~\cite{mu_dilemma}. Due to the smaller system sizes and the larger 
trap curvatures than in 2D, the twenty-four nearest neighbors lattices 
sites could not be considered since the effective chemical potentials at 
these nearest neighbors lattice sites and at the central lattice sites 
differ significantly from each other and thus correspond to two different 
scan lines in the phase diagram. As can be extracted from 
Fig.~\ref{fig:3D_1}, the same yet less pronounced behavior in regards of 
bulk quantum criticality is observed, i.e., locally at the center of the traps, 
the transition from the local superfluid to the local Mott-insulating 
state is found to have slightly stronger critical fluctuations
in the quartic than in the 
quadratic trap. However, care must be taken while interpreting these 
results, due to the finite size effects that strongly affect the centers 
of these small 3D systems with their significantly reduced linear extents, 
compared to the 2D systems presented above.

\section{Conclusions}
\label{sec:conclusions}

Ultracold bosonic atoms in optical lattices have been studied in quadratic 
and quartic traps.  The physics of these confined systems is almost 
perfectly captured by the Bose-Hubbard model~\cite{jaksch:98}. The 
specific traps can be taken into account by a site-dependent effective 
chemical potential in the Bose-Hubbard Hamiltonian. Our results in 2D 
clearly demonstrate that the quantum criticality of the transition 
at the surface layer separating the coexisting superfluid and Mott-insulating 
states is destroyed by the finite gradient of the trapping potential, 
even for traps with quartic profiles, which have steep walls and in whose 
centers the gradient almost vanishes. 
However, we have found that, if the measurements are restricted to the 
center of the traps and the local state is driven through the 
superfluid--Mott insulator transition, the then observed bulk transition 
has stronger critical fluctuations in quartic traps than in quadratic traps. 
To demonstrate this, 
we use the local compressibility as a genuine, local order parameter to 
characterize the transition and drive the system center from the 
superfluid to the Mott-insulating state by decreasing the ratio $t/U$ 
between the hopping and the interaction parameter at constant chemical 
potential $\mu/U$. The deviations of the curves obtained from the systems 
with quartic traps from the curves obtained from uniform systems with 
closed boxes and periodic boundary conditions are much less pronounced 
than the deviations of the curves obtained from the systems with quadratic 
traps. Even if there is no perfect match, we expect the quartic traps to 
constitute a remarkably better prerequisite for the experimental detection 
of true quantum criticality.

Our simulations of moderately-sized 3D systems lead to the same 
conclusions as in 2D, but, due to strong finite size effects 
present in these systems, an analysis with yet larger systems is desirable.

Finally, it is worth mentioning that quartic traps represent an ideal for 
traps that are experimentally realized by superimposing blue- to 
red-detuned laser beams with Gaussian intensity profiles. The flatness 
achievable in the trap center depends on the ability to suppress the 
quadratic terms in the series expansion of the resulting confining 
potential. Since this might turn out to be a rather difficult task, it 
could be interesting to investigate what influence small fluctuations 
in the otherwise flat trap centers have on the local bulk quantum criticality.

\begin{acknowledgments}
The simulations have been performed on the Hreidar and Gonzales clusters 
at ETH Z\"urich using the worm algorithm~\cite{prokof'ev:98} and the SSE 
algorithm with directed loops~\cite{alet:05a} of the ALPS~\cite{alet:05b} 
project. We thank T.~Esslinger, M.~K\"ohl, and S.~Trebst for helpful 
discussions.
\end{acknowledgments}

\bibliography{comment,refs}

\begin{thebibliography}{20}
\bibitem[{mu_({\natexlab{a}})}]{mu_scan}
\bibinfo{note}{For example, in the constant $\mu/U$ scan presented in
  Fig.~\ref{fig:phase_diagram}, the filling of the uniform system along the
  part of the scan that lies in first Mott lobe is necessarily commensurate,
  whereas the filling along the part of the scan that lies in the surrounding
  superfluid region is in general incommensurate. Furthermore, for the
  simulations at constant $\mu/U$ of the systems with quadratic and quartic
  traps, the particle number slightly depends on $t/U$.}
\bibitem[{mu_({\natexlab{b}})}]{mu_dilemma}
\bibinfo{note}{Note that, since all our 2D and 3D systems have an even number
  of lattice sites, there is no lattice site at the center of the system, at
  which the observables can be measured for the constant $(\mu/U)$ scans. The
  observables are thus measured at the closest lattice sites that surround the
  center. For each system with quadratic or quartic traps, the values of the
  chemical potential offsets $\mu_0/U$ have been chosen in such a way that the
  effective chemical potential $\mu^{\mathrm{eff}}_{\mathrm{center}}/U$ at
  these central lattice sites coincides with the chemical potential
  $\mu/U=0.217$ chosen for the uniform systems.}
\end{thebibliography}

\end{document}